\documentstyle{EuroPhys}

\input EuroMacr
\input epsf

\begin{document}

\shorttitle{A microscopic model for solidification} 
\title{A microscopic model for solidification}
\author{M. Conti$^1$, U. Marini Bettolo Marconi$^1$ and A. Crisanti$^2$}

\institute{$^1$Dipartimento di Matematica e Fisica, Universit\`a di
Camerino and Istituto Nazionale di Fisica della Materia,
Unit\`a di Camerino,
Via Madonna delle Carceri, I-62032, Camerino, Italy,
$^2$ Dip. di Fisica, La Sapienza, Roma ,Italy}
\rec{ }{ }
\pacs{
\Pacs{64}{60.-i}{General studies of phase transitions}
\Pacs{64}{60.Cn}{Order disorder transformations}
}
\maketitle
\begin{abstract}
 We present a novel picture of a non isothermal solidification process
starting from a molecular level, where the microscopic origin of the 
basic mechanisms and of the instabilities 
characterizing the approach to equilibrium 
is rendered more apparent than in existing approaches based 
on coarse grained free energy functionals \`a la Landau. 

The system is composed by a lattice of Potts spins, which
change their state according to the stochastic dynamics proposed some time ago
by Creutz. Such a method is extended to include the presence
of latent heat and thermal conduction. 

Not only the model agrees with previous continuum treatments,
but it allows to introduce in a consistent fashion 
the microscopic stochastic fluctuations. These play an
important role in nucleating the growing solid phase in the melt.
The approach is also very satisfactory from the quantitative 
point of view since the relevant growth
regimes are fully characterized in terms of scaling exponents.
 
\end{abstract}
\vspace{0.2cm}

\vskip2pc

 A  remarkable feature, observed in solidification
processes, is the occurrence
of different morphologies of the solid-liquid boundary, namely 
flat, cellular or dendritic, according
to the initial undercooling, which acts as control 
parameter for the growth process.

The theoretical models employed so far to study this kind of
non equilibrium phase transformations represent at 
different degrees a coarse grained, mesoscopic
picture of the underlying microscopic processes. They
miss consequently  an important point, i.e. the stochastic
character of the thermal fluctuations and
disregard the nucleation of the growing phase in the
melt. It has also been argued \cite{karma} that small a
thermal noise  can account for the enhanced sidebranch activity
during dendritic growth.

 The state of affairs is as follows: 
in the free boundary model \cite{freeboundary}
\cite{Langer}  the solidification 
process is formulated
in terms of a sharp moving boundary, acting as a 
source for the diffusion field, while the thermodynamic 
information enters the model only via the boundary conditions;
 in the phase field model \cite{phasefield}, instead, employing a local 
order parameter formulation, whose evolution depends on some 
coarse grained Ginzburg-Landau functional,
one regards 
the transition region between the two phases as smooth
and characterized by arbitrary topology. In spite of its merits the latter
approach is frankly mesoscopic and the temperature acts
only as a parameter which controls the relative stability of the two phases,
but does not play a role in the nucleation processes.

In the present paper we shall adopt a molecular point of view,
instead of a continuum formulation. In 
spite of the fact that in many cases of interest, 
neither of the
methods can provide information which is not also accessible to the other,
 the microscopic approach is required for the study of
phenomena involving large spatial gradients and is able to account 
for noise effects. In the following we shall 
discuss how to incorporate in a consistent
fashion the stochastic character of fluctuations into a microscopic model 
of solidification \cite{karma}.

The formulation of a minimal model to explain the
occurrence of the instabilities and the variety of patterns during 
the crystal growth from the melt needs to include four basic effects:

i) the tendency of the molecules towards an ordered state
which minimizes their configurational energy. 

ii) the opposite tendency of the system towards a disordered state
which  maximizes its entropy.

iii) the release of latent heat during the liquid/solid transformation

iv) the diffusion of such a heat away from the two phase boundaries.

To fulfil the above requirements and maintain a microscopic point of view
we shall consider a novel lattice model 
by assigning to each site $i$ a 
Potts spin variable which takes on the integer values $1\leq n_i \leq q$.
When $n_i=1$, the spin energy is negative and
given by $-\lambda<0$, otherwise it
is zero. An interface between a  spin in a state $1$ and a spin at 
a nearest neighbor site $j$ whose
state is $n_j>1$ has a surface cost  $J>0$. 
At equilibrium and in the thermodynamic 
limit such a model displays two macroscopic phases, characterized by an order 
parameter $m=(q<\delta_{1,n}>-1)/(q-1)$.
The ordered phase $m\simeq 1$,
and the disordered phase constituted by a mixture of the
$(q-1)$ microscopic states with $n_i>1$ coexist at  
$T_m\simeq\lambda/\ln(q-1)$. Beneath $T_m$, the melting temperature, 
the ordered
phase is absolutely stable, whereas above $T_m$ the 
entropic phase $m\simeq 0$ has a lower free energy.
 The competition between the large Boltzmann weight of the 
state with $n=1$ and the larger number of microscopic configurations 
of the disordered state determines the dominance of one
phase over the other. The analysis of the free energy gives a
complete description of the equilibrium properties of the system.
 Our interest, however is to study the evolution towards equilibrium, 
when the system is prepared in a metastable initial state.

To achieve that we introduce a dynamics by adding $p$ auxiliary 
degrees of freedom
at each site, the 
Creutz demons \cite{creutz}, characterized by energies $\epsilon^{\alpha}_i$
(with $1 \leq \alpha\leq p$) which act as energy reservoirs and 
control the evolution of the spin subsystem. When
a variable changes from a value $n_i$ to $n_i'$, 
the demon energy varies from
$\epsilon^{\alpha}_i$ to $\epsilon^{\alpha}_i-\Delta \epsilon$,
having chosen $\alpha$ randomly among the $p$ possible
values and where $\Delta \epsilon$ represents the spin energy 
change. Attempted moves which render negative $\epsilon^{\alpha}_i$
are rejected. 
 To complete the model we must include a
heat conduction mechanism by assuming that at every
time step an 
elastic collision event takes place with probability $\kappa$ between the 
nearest neighbors demons $i$ and $j$
according the rule \cite{ulam} that the postcollisional energies are
$\epsilon^{\alpha}_i= (\epsilon^{\alpha}_i+\epsilon^{\beta}_j)r$ and
$\epsilon^{\beta}_j= (\epsilon^{\alpha}_i+\epsilon^{\beta}_j)(1-r)$
where $r$ is a random number
chosen from a uniform distribution in the unit interval.
 Physically one can think of the demons as kinetic degrees of freedom,
irrelevant, as far as equilibrium configurational properties
are involved, but necessary to mediate the energy exchanges
during the approach to equilibrium.
Since
the temperature may have spatio-temporal fluctuations,
one needs to consider a local temperature given by the 
average  energy of a demon over a small time interval. 
It can be showin redistributing the energy, i.e.
in conducting the heat
\cite{ulam}, even in the absence of spin dynamics \cite{grant}.
 It is interesting to stress that 
such a mechanism allows to obtain the correct Maxwell-Boltzmann
distribution and that with different rules
of energy redistribution  such a behaviour is not recovered.

Thus sweeping randomly the lattice  
and updating sequentially  the spins and the demons 
the system behaves ergodically and 
one can compute meaningful statistical averages.
With the set of rules stated above we are able to simulate a system
which evolves in adiabatic conditions, i.e.  at constant energy, but at
varying temperature.  This approach can be employed in many cases where
the more traditional isothermal Kawasaki and Monte Carlo methods 
\cite{simulmethod} do
not represent faithful descriptions of the transients encountered in
experimental situations.  As we shall see below we are able to capture
many features relevant in solidification processes. Of course, after 
an appropriate thermalization run the demon energy distribution becomes
statistically uniform, i.e. the temperature attains  a uniform
value, and the algorithm
becomes equivalent to the Metropolis Monte Carlo method. 

 As a first check we verified
that a two dimensional system of demons, with fixed temperatures at the
top and bottom boundaries, reaches a steady state in which the
temperature profile is linear from the cold to the hot wall and the
thermal conductivity is independent of the temperature.  To tune the thermal
conductivity we allowed only a randomly 
chosen fraction $\kappa$ of the demons to undergo the
collisional dynamics.

Instead, in order to recover 
the standard picture of the solidification 
we consider a  demon with low energy next to a site in 
state $1$; it can be seen that such a demon favors
the transformation of liquid into solid; as a consequence the 
interface advances
and the demon energy increases. But since the energy produced 
by the phase change must be diffused away
before new solid can be formed, the process is slowed down;
the ordered phase may also form protrusions into the colder
phase in order to diffuse away more efficiently the latent heat,
maximizing its surface.  An appropriate choice of ratio bulk-gain versus
surface-cost $\lambda/J$ renders the system stable with respect
to the formation of solid islands within the liquid phase, and the
growth proceeds only at the interphase boundaries.
Conversely, if the demon energy is high 
all the states become equally accessible, and there is no tendency to 
crystallize.

Let us turn to the quantitative predictions of the model 
and give
numerical evidence on one and two dimensional
lattices.  The most relevant
macroscopic control parameter in determining the dynamics of the process
is the dimensionless undercooling $\Delta$, which in terms of the specific
heat $C$ and of the latent heat of fusion $L$ can be written as
$\Delta=C(T_m-T_0)/L$, where $T_m$ and $T_0$ represent the coexistence
temperature of the two phases and the initial temperature of the melt,
respectively.  As we chose in all the simulations melting temperatures
well below the temperature $T_t$,
where the solid metastable branch shows a peak in the
specific heat, for all
practical purposes the only significant contribution to 
the specific heat is due to
the $p$ degrees of freedom of the demons, so that $\Delta$ can be
expressed in terms of our microscopic variables as: 
$\Delta=p[T_m - T_0]/\lambda $. Notice that such a formula
represents an important link between the microscopic
parameters $p$ and $\Delta$ and the macroscopic 
thermodynamic quantities $C$ and $L$.

Since $T_m$ is proportional to $\lambda$, it appears clear the
physical motivation of having more than one 
demon per site, i.e. a large specific heat is needed
in order to attain large undercoolings.
In other words one can 'solidify' an arbitrary amount of material
by being able to adsorb through the demons the latent heat released.

In one dimension a system with $L_x$ lattice sites aligned along the $x$
axis was prepared in a nonequilibrium initial state, with an interface
boundary at $x=x_0$ separating a solid phase ($n_i=1$) and a liquid
phase in which $n_i$ was randomly distributed in the range $2\leq n_i
\leq q$.  The average demon energy was set to $T_m$ in the solid and to
$T_0<T_m$ in the liquid.  Then the evolution of the system was followed
up to $N$ Monte Carlo steps per site (MCS), until a well identified
growth regime was attained.  For each simulation an ensemble average
over 256 different runs was taken to get statistical significance.  A
value $L_x=1400$ was sufficient to avoid finite size effects in all the
simulations.  To focus on the interface dynamics, and to prevent
nucleation in the bulk liquid, we chose a high value of the interface
energy cost ($J=4$).  Notice that in one dimension $J$ has a strong
effect on the nucleation rate in the liquid but does not affect the
interfacial growth rate, which depends solely on $\Delta$ . 
Let us see how, in one dimension, 
the choice of $\Delta$ determines 
different growth regimes of the dynamical solutions of the
model.

In fig. \ref{fig:fig1} the dashed lines show two different scaling laws:
the upper one displays the interface position versus time
for $\Delta = 1.2$,
$\lambda =1$, $p=4$, $q=20$ and the thermal conductivity parameter
$\kappa=0.08$.  The long time behaviour of the interface motion
clearly converges to a steady solution with constant velocity,
in good agreement with the macroscopic 
free boundary model \cite{freeboundary}.
The lower dashes instead, 
refer to the same set of 
parameters, but  $\Delta=0.8$; and                                      
the interface velocity decays as $v \sim t^{-1/2}$.
We verified that, in general, when $\Delta>1$                               
the front travels at constant velocity $v \propto (\Delta -1)$, but shows a
significant deviation 
from such a formula near the crossover point to the diffusive regime,
$\Delta=1$.

 Even more interestingly and more conclusively,
we observe that the temperature at the interface, $T_i$,
as measured through the demon energy distribution, is lower than the 
equilibrium melting temperature by a quantity which is proportional 
to the interfacial velocity $v$.
This is the so called  kinetic undercooling of the 
solid-liquid interface, 
which in the free boundary model must be introduced ad hoc
through the constitutive law $T_i = T_m - \beta v$, to
account for the necessity of a departure from local thermodynamic
equilibrium (i.e.  of a finite free energy difference between the two
phases) to advance the solidification process.
We 
evidenced the presence of kinetic undercooling at a microscopic level
through a series of simulations performed with $\lambda
=1$, $p=2$, $q=4$ and $\kappa=0.6$.  The initial undercooling $\Delta$ ranged
from $1.12 \leq \Delta \leq 1.82$.  For each simulation, in the steady
regime we evaluated the interface temperature $T_i$ through an ensemble
average of the energy of the demons located at the solid-liquid
boundary.  The results are shown in fig.
\ref{fig:fig2}. One can see that the
driving force for the solidification process, i.e.  the shift $T_m-T_i$
is an increasing function of the growth rate; according to the
predictions of both the free boundary and the phase field models the
dependence ${T_i}(v)$ is found to be linear to a quite good extent.  It
is worth noting that the solid straight line, which represents the best
fit of the data points, intersects the vertical axis at a temperature
$T=0.87$, very close to the equilibrium melting temperature $T_m\sim
0.91$.

As we consider the two dimensional case 
we see that the solid-liquid interface undergoes a morphological
instability due to the competition between two effects.  The necessity
to diffuse the latent heat away from the interface favors the formation
of protrusions of the growing solid into the supercooled melt; on the
other hand the surface tension $\sigma$, via the Gibbs-Thomson
effect, tends to restore the minimum surface configuration \cite{gibbs}.  
The characteristic length scale for the resulting pattern has been
identified by Mullins and Sekerka 
\cite{sekerka} as $\lambda_0 \sim 2\pi
{(3{d_0}{\ell})}^{(1/2)}$, where $d_0$ is the capillary length, defined
as $d_0={\sigma T_m C}/{L^2}$ and $\ell$ is the thermal diffusion
length.  For steady growth we have $\ell = D/v$, where $D$ is the
thermal diffusivity of the melt.
We considered a rectangular domain $0\leq x \leq L_x$, $0\leq y \leq L_y$ with
$L_x = 128$ and $L_y = 256$.  The system was prepared with a solid
phase ($x<10$) and a liquid phase ($x>10$) separated by an interface
parallel to the $y$ axis; the demons energies were distributed to fix
the initial temperature in the solid at $T_m$; the liquid was
undercooled at $\Delta=0.6$.  The other parameters of the model were
selected as $q=20$, $p=3$, $\lambda = 1$, $J=0.24$, $\kappa=0.02$.  With
this choice the growth process proceeded essentially through activated
nucleations at the solid-liquid interface, while nucleation in the bulk
liquid was almost suppressed.  We checked also that the far field
temperature was never affected by the finite size of the system.  The upper
portion of Fig.
\ref{fig:fig3} shows the growing solid after $24 \cdot 10^3$ MCS.  
The dendritic
structure of the pattern is clearly visible.  An analysis of the
temperature field indicates a value of the thermal diffusion
length $\ell \sim 35$ lattice sites.  As the surface tension is of the
order of $J$, the characteristic width of the fingers should be of the
order of $\sim 30$ lattice sites, in good agreement with the numerical
results.

Recent advances on pattern formation in diffusional fields indicate that
no stable dendritic fingers can grow, unless a preferred direction is
injected into the model through anisotropy of the surface tension or
geometrical constraints
such as the directional solidification
in a channel \cite{marinozzi}.  In order to render the surface tension 
anisotropic we differentiated the energy cost $J_n$ between
the nearest neighbours and the
next nearest neighbours $J_{nn}$.  
The lower part of Fig. \ref{fig:fig3}
shows the dendritic pattern obtained with the same choice of
parameters as above, and with $J_n=0.36$,$J_{nn}=0.12$.  The
directional growth of the fingers along the channel axis is now better
established; the tip splitting dynamics and the competition between
the different fingers is clearly recognizable.

 Before concluding we notice that in the present approach the
heat released to produce a solid particle from the melt increases
the demon energy and the local temperature. The latter in turn,
according to the  
Arrhenius relation \cite{Langer2} increases the rate of nucleation 
$\tau^{-1} \simeq \exp(-\Delta F/k_B T)$, where  
$\Delta F$ represents the typical activation barrier  
to be overcome in order to crystallize.  We observe that the 
liquid phase can be appreciably undercooled below $T_m$, but the solid 
cannot be overheated at the same extent ; in other words
we are able to eliminate
the spurious symmetry between the two processes, which is
present in the Phase Field treatment \cite{Conti}
-\cite{iori}, where the temperature 
merely determines the relative stability of the two phases and the driving
force towards equilibrium, but has no influence on the nucleation.
 Such a non linear feedback mechanism is absent in all previous approaches
to the problem and is one of the successes of the present model.

To summarize in the present letter we have introduced a new lattice model to
simulate the microscopic behavior of a system undergoing a first
order transition with emission of latent heat. The dynamics we have
proposed 
does not need fine tuning of parameters
and differs from the existing stochastic approaches 
because the temperature varies in space and in time. 
We avoid the traditional difficulty with the Phase Field model
of treating the temperature as an external field which controls the
relative height of the two minima of the free energy, i.e. the difference
in free energy between the liquid and the solid phase and not including
noise terms.  Doing so we can describe 
on equal footing the microscopic fluctuations which make possible the
nucleation of new islands of solid and the field which drives the growth.

 With respect to earlier work \cite{grant}
we observe that in a different model for solidification,
based on a lattice approach,
two important differences could be devised:
i) the specific heat is only due to the spins
and strongly temperature dependent, while in
our case is proportional to the number, p, of demons
and therefore mimics the Dulong-Petit law. 
ii) the thermal conductivity is associated only with
the spin flip mechanism and vanishes in the infinite temperature limit
while at low temperature becomes a thermally activated 
process, while in our work is nearly temperature independent.

Concerning future perspectives it remains to investigate to
which macroscopic description our microscopic model corresponds;
this can be achieved by a suitable coarse graining procedure. 
Besides one could apply the present approach
to conserved order parameter dynamics, such as melting of binary alloys,
or study systems quenched from a high temperature to a low
temperature condition employing a finite cooling rate.
 Finally we mention the possibility of considering different 
energy redistribution laws among demons, i.e. different kind of noises.

\newpage

\begin{figure}[h]
\caption{ Upper and lower dashes represent the average interface position 
in lattice units
in the kinetic regime ($\Delta=1.2$) and in the diffusive regime
($\Delta=0.8$), respectively.
The solid lines represent the power laws $\sim t$ (upper) and 
$\sim t^{1/2}$ (lower), respectively. Time units are in MCS.} 
\label{fig:fig1}
\end{figure}

\begin{figure}[h]
\caption{ The dots represent the interface temperature (in energy units)
versus the interface velocity (lattice sites
per MCS); the solid line is drawn as guide
to the eye.}
\label{fig:fig2}
\end{figure}

\begin{figure}[h]
\caption{ Typical growth patterns with anisotropy (bottom) and
without (top). Notice the bulk nucleation away form the boundaries. 
The parameters are given in the text.}
\label{fig:fig3}
\end{figure}


\begin{thebibliography}{99}
\bibitem{gibbs}
Within the model solid droplets immersed
in a sea of liquid display the Gibbs-Thomson effect, i.e. the 
lowering of the melting temperature due to the curvature. An analytical 
estimate of such shift is:
$
T(R)=T_m-J/[R \ln(q-1)]
$
where $R$ is the droplet radius.
\end{thebibliography}
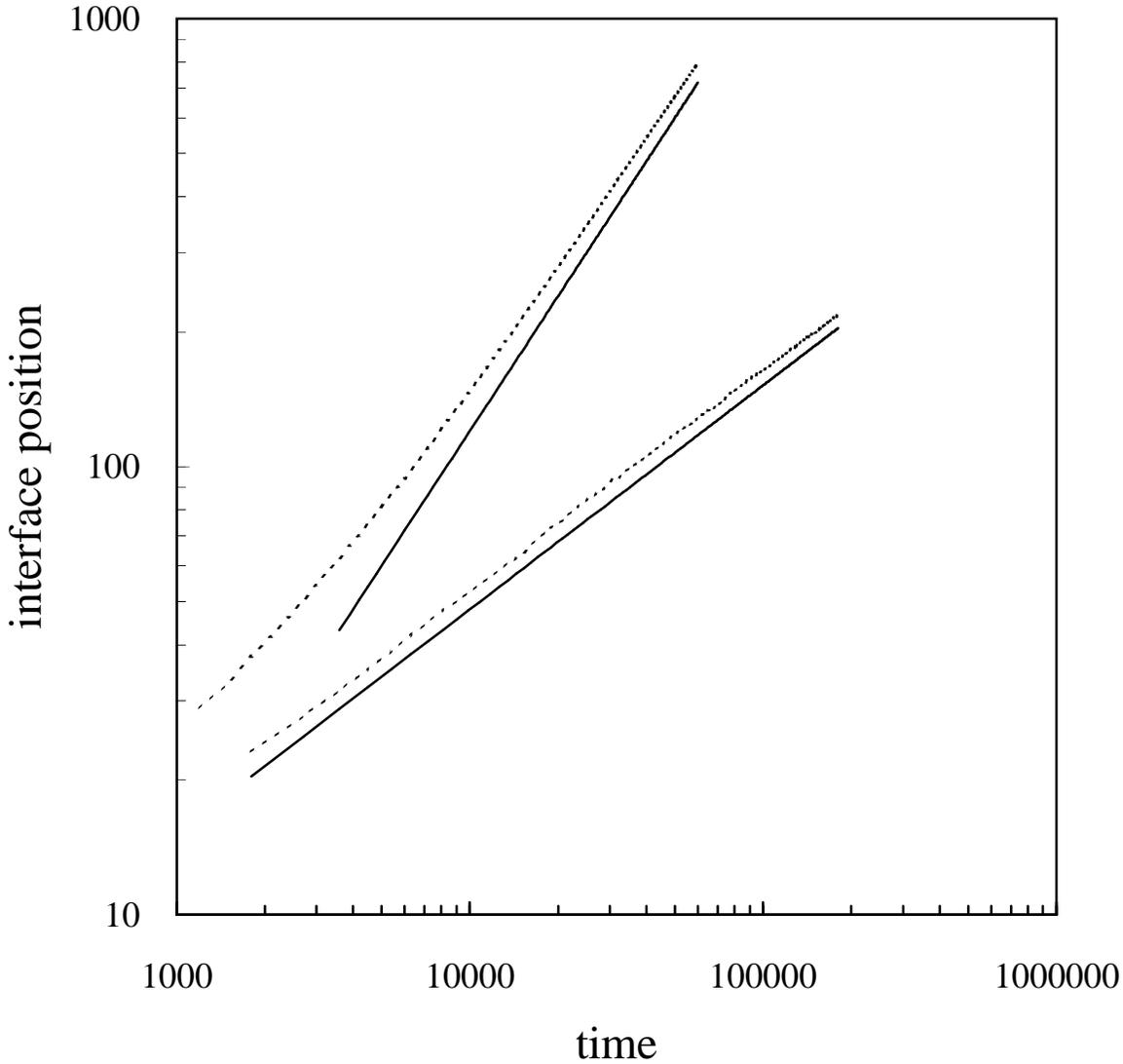
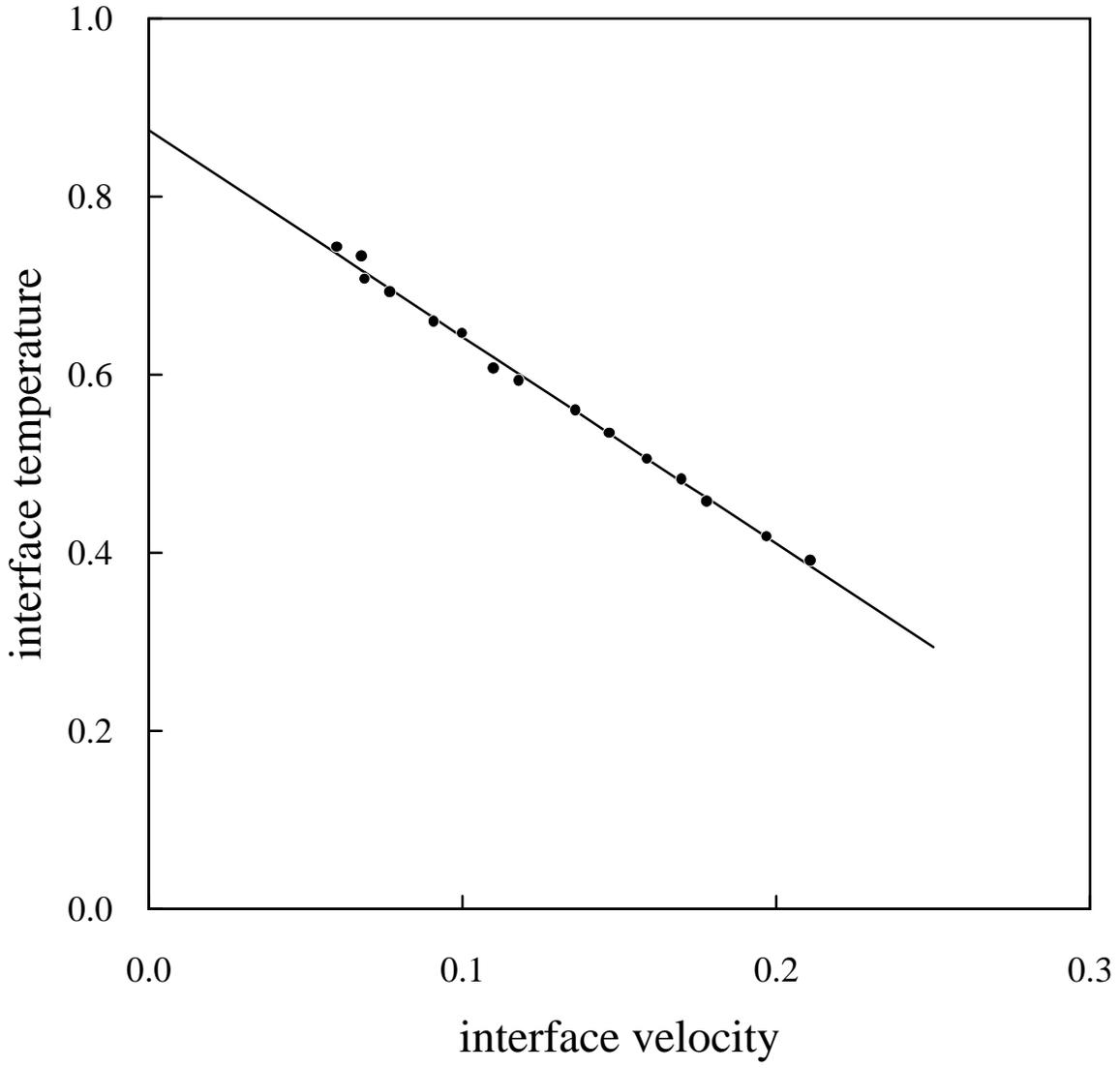
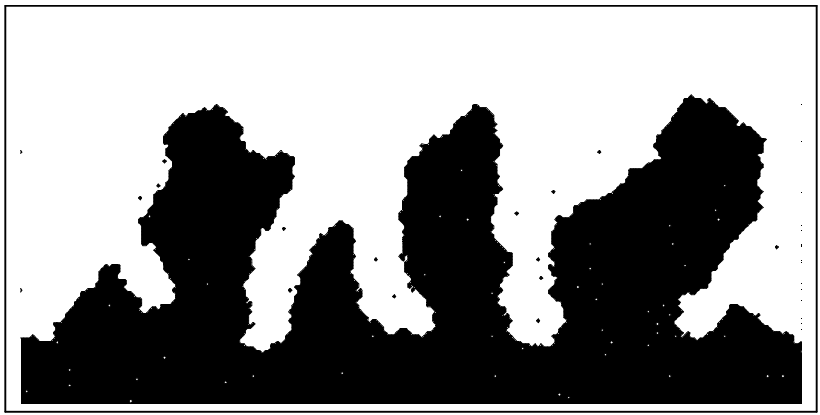
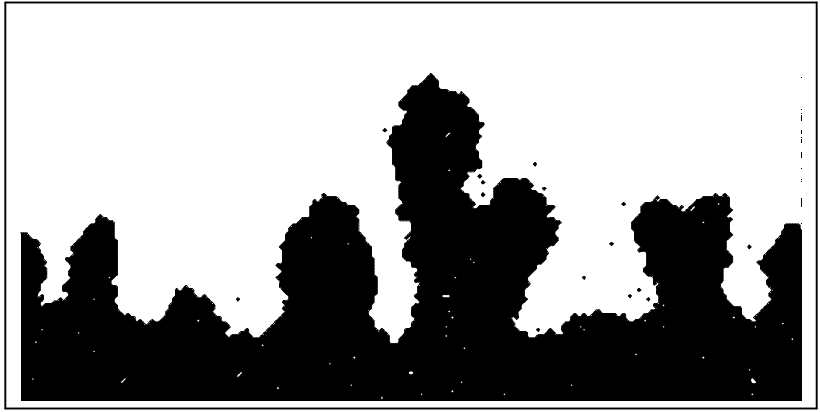
\end{document}